\documentstyle[pre,twocolumn,aps,epsf,floats]{revtex}

\begin{document}
\draft
\title{Synchronization, Diversity, and Topology of Networks of
Integrate and Fire Oscillators.}

\author{X. Guardiola, A. D\'{\i}az-Guilera, M. Llas and C. J. P\'{e}rez}
\address{
Departament de F\'{\i}sica Fonamental, Universitat de
 Barcelona, Diagonal 647, E-08028
Barcelona, Spain \\}

\date{\today}
\maketitle
\begin{abstract}
We study synchronization dynamics of a population of pulse-coupled
oscillators. In particular, we focus our attention in the
interplay between networks topological disorder and its
synchronization features. Firstly, we analyze synchronization time
$T$ in random networks, and find a scaling law which relates $T$
to networks connectivity. Then, we carry on comparing
synchronization time for several other topological configurations,
characterized by a different degree of randomness. The analysis
shows that regular lattices perform better than any other
disordered network. The fact can be understood by considering the
variability in the number of links between two adjacent neighbors.
This phenomenon is equivalent to have a non-random topology with a
distribution of interactions and it can be removed by an adequate
local normalization of the couplings.
\end{abstract}


\section{Introduction}
Synchronization of populations of interacting oscillatory units
takes place in several physical, chemical, biological and even
social systems\cite{Kuramoto,Winfree,clap}. Networks of
interacting oscillators are currently used to modelize these
phenomena. In this paper we will focus on a special kind of
interacting oscillators, namely pulse coupled oscillators. These
units oscillate periodically in time and interact, each time they
complete an oscillation,  with its coupled neighbors sending them
pulses which modify their current states. These systems show a
rich spectrum of possible behaviors which ranges from global
synchronization\cite{Strogatz} or spatio-temporal pattern
formation\cite{Bressloff,Me} to self-organized
criticality\cite{SOC}. Although some theoretical approaches have
been proposed, in general, the singular nature of pulse-like
interactions does not allow to describe the system in terms of
tractable differential equations. Despite this, some methods have
been developed to find the attractors of the dynamics and study
their relative stability\cite{Strogatz,PRE,Bressloff}.

In this paper, we want to focus on the effects that different
topologies have on the dynamical properties of the network. In
particular, we will study how does network's topology affect
global synchronization. So far, most of the studies on networks of
coupled oscillators have been done on either small connectivity
lattices (usually 1D rings)\cite{PRE,Bressloff} or globally
coupled networks (all to all coupling)\cite{Strogatz}.
Nevertheless, there is some work done in networks of continuously
coupled oscillators (Kuramoto's)\cite{Niebur,Huberman,SW} and
Hodgkin-Huxley neuron-like models\cite{Huerta} where different
non-standard topologies are considered. Among these, the so-called
{\em Small-world} networks seem to be an optimal arquitecture, in
terms of activity coherence, for some of these coupled
systems\cite{SW,Nature}.

Pulse coupled oscillators are commonly used to modelize driven
biological units such as pacemaker cells of the heart\cite{Peskin}
and some types of neurons\cite{Torras}. In these systems,
synchronization is usually considered to be a relevant state.
Regarding heart, pacemakers must be synchronized in order to give
the correct heart rhythm avoiding arrhythmias or other perturbed
states. In populations of neurons, synchronization has been
experimentally reported\cite{Gray} and is believed to play a role
in information codification\cite{Sompolinsky}. Therefore, it is
interesting to check which kind of topologies makes the network
reach a coherent state more easily and uncover why is it so by
looking for its responsible mechanisms. We will focus in a whole
family of networks which are characterized by its increasing
degree of disorder, i.e. ranging from regular lattices to
completely random networks.

The structure of this paper is the following. In Sec. II we
introduce the model of pulse-coupled oscillators which is going to
be used throughout the paper. In Sec. III we start studying
synchronization of populations of these coupled oscillators in
random networks. In Sec. IV we compare random network performance
with the more classical regular lattices. In Sec. V we consider a
more general family of networks with a variable degree of
randomness and study its synchronization properties. Moreover, the
interplay between diversity, interaction and topology is also
discussed. In the final Section we present our conclusions.

\section{Basics}
We study the synchronization of a network of $N$ oscillators
interacting via pulses. The phase of each oscillator $\phi_i$
evolves linearly in time

\begin{equation}
\frac{d\phi_i}{dt}=1 \hspace{2em}\forall i=1,\ldots ,N
\end{equation}
until one of them reaches the threshold value $\phi_{th}=1$. When
this happens the oscillator fires and changes the state of all its
vicinity  according to
\[
\phi_{i}\geq 1 \Rightarrow \left\{
\begin{array}{l}
\phi_{i}\rightarrow 0 \\
\phi_{j}\rightarrow\phi_{j}+ \Delta(\phi_{j})
\end{array}
\right.
,
\]
where $j \subset \Gamma(i)$, $\Gamma(i)$ being the list of nearest
neighbors of oscillator $i$. The nonlinear interaction is
introduced in the {\em Phase Response Curve} (PRC) $\Delta(\phi)$.
We use a PRC which induces a global synchronization
$(\phi_1=\phi_2=...=\phi_N)$ of the population of oscillators:
$\Delta(\phi)=\varepsilon\phi$ with $\varepsilon>0$ . This PRC is,
indeed, the most simple type of interaction that always leads to a
synchronizated state whatever the initial conditions are. In other
words, synchronization is the unique attractor of the dynamics.
Altough it has only been mathematically proved for
all-to-all\cite{Strogatz} and local\cite{PRE} couplings, for all
the topologies we have dealt with, synchronization holds.
Therefore the dynamics could also be expressed as

\begin{equation}
\frac{d\phi_i}{dt}=1+\varepsilon\sum_{j \subset \Gamma(i)}\delta(t-t_j),
\end{equation}
where $t_j$ are the firing times of $\phi_j$. To define a certain
degree of synchronization in our simulations, we define the
variable

\begin{equation}
m \equiv \frac{1}{N}\sum_{i=1}^{N}(1-\phi_{i}(t_1^+))
\end{equation}
measured each time $\phi_1=0$. The choice of oscillator $1$ as a
reference is completely arbitrary. Notice that measuring the
phases at $t_1^+$ ensures that these phases are $0$ if they are
synchronized with oscillator $1$, not depending on the order they
fire. In this way, we have a series of system ``snap-shots'' which
mathematically correspond to a {\em return map} of the dynamics
(see Fig. 1). {\em Synchronization time} $T$ is thus defined as
the time needed to reach $m=1$. When this happens, all oscillators
will always fire in unison.

\begin{figure}
\label{transit}
\centerline{
        \epsfxsize= 6.0cm
        \epsffile{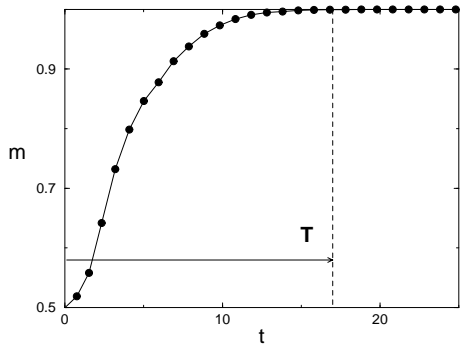}}
        \caption{ Typical evolution of $m$. The activity
        coherence of the system increases with time. At time
        $T$ the population reaches global
        synchronization ($m(t=T)=1$). In this example, we
        consider a population of $N=300$ oscillators with arbitrary
        initial conditions ($m(t=0)\sim 0.5$).}
\end{figure}

\section{Random networks}
We start studying synchronization of a population of coupled
oscillators by defining a Random Network ($RN$). We restrict
ourselves to the most simple type of $RN$ \cite{erdos}, that is,
we randomly select a pair of the $N$ nodes and establish a link
among them, repeating this procedure up to a certain number of
links $l$. Notice that, with this wiring method, there is no
guarantee of ending up with a connected network (where there must
be a path connecting {\em any} pair of nodes). Dealing with a
network splitted into two or more clusters would make global
synchronization ($m=1$) impossible so we should avoid such
pathological configurations. In order to have a connected network
one has to work over a threshold number of links which ensures
connectivity \cite{ballobas}

\begin{equation}
l\gg\frac{N}{2}\ln(N).
\end{equation}
Therefore we can study RN whose number of links $l$ runs from the
above limit up to the globally connected network (all to all
coupling) which has $l=N(N-1)/2$. In addition, if we want to study
the transient to synchronization, one should always stay in the
limit

\begin{equation}
N\varepsilon \ll 1,
\end{equation}
otherwise synchronization would be achieved in a few firings due
to the stronger interaction.

As one expects, when the number of links $l$ is increased, the
time needed to reach synchronization $T$ diminishes, having its
lowest value for the globally connected case. What is really
interesting is {\em how} does $T$ decrease as we consider networks
with more links. It turns out that $T$ follows a power-law with a
slope which is independent of the number of nodes.

\begin{equation}
T\sim l^{-\alpha} \hspace{2em} \forall N,
\end{equation}
with $\alpha=1.30\pm0.05$ for $\varepsilon=0.01$. In Fig. 2 this
behavior is shown. In these simulation results, each point is
averaged over different random topologies with the same number of
links and over different initial arbitrary conditions for all the
oscillators.

\begin{figure}
\label{pendents}
\centerline{
        \epsfxsize= 8.0cm
        \epsffile{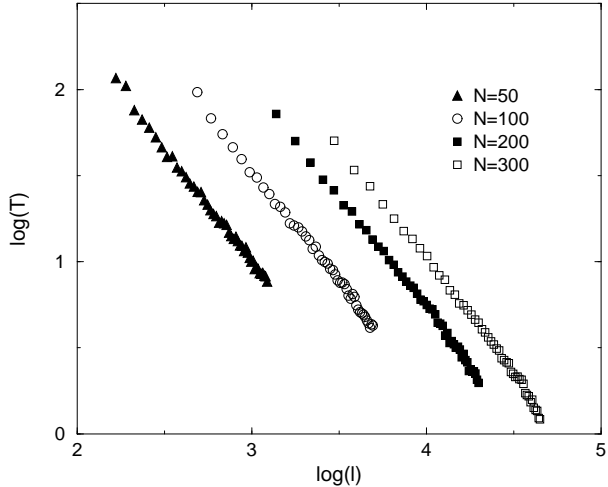}}
        \caption{$T$ as a function of $l$ for a fixed number of
        oscillators $N$ and fixed interaction strength
        $\varepsilon=0.01$. Results are averaged over 100
        different arbitrary initial conditions in 10 different
        random topologies. The lowest value of $l$ is in each
        case the one which statistically guarantees that the network
        is connected and the highest one is all-to-all coupling.
        For higher values of $N$ the power-law behavior is lost
        since relation (3) does not hold anymore.}
\end{figure}
In addition, one can study how does $T$ increase with the number
of oscillators $N$. We find that it also follows a power-law
behaviour which does not depend on the number of links $l$
considered

\begin{equation}
T\sim N^{\beta} \hspace{2em} \forall l,
\end{equation}
with $\beta=1.50\pm0.05$ for $\varepsilon=0.01$. Therefore, once
the interaction strength is set, we can characterize
synchronization time $T$ by means of network's geometrical
properties trough the scaling relation

\begin{equation}
T \sim \frac{N^\beta}{l^\alpha}.
\end{equation}
which can be rewritten as
\begin{equation}
\frac{T}{N^{2\alpha-\beta}}\sim \left(\frac{l}{N^2}\right)^{\alpha},
\end{equation}
In Fig. 3 we plot the collapse of data curves according to Eq. 9
and the agreement is excellent. The exponents $\alpha$ and $\beta$
are constant within the error bars for the checked values of
$\varepsilon$ ($0.1>\varepsilon >0.005$).

\begin{figure}
\label{colapse}
\centerline{
        \epsfxsize= 8.0cm
        \epsffile{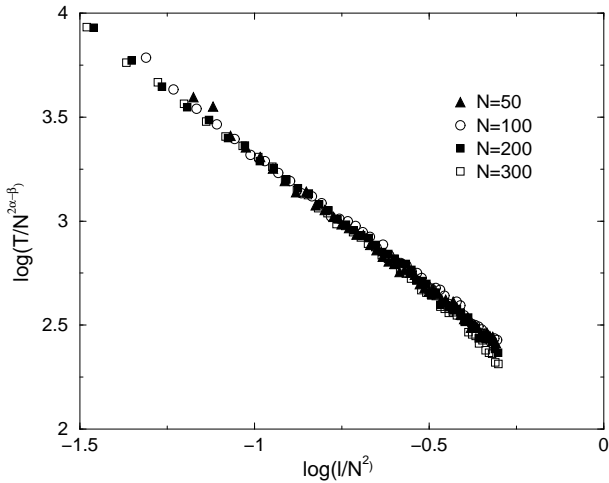}}
        \caption{Collapse of data curves $T(l,N)$ supporting the
        scaling hypothesis $T \sim N^\beta l^{-\alpha}$
        for $\varepsilon=0.01$.}
\end{figure}

\section{Random Networks versus Regular Lattices}

Once we have seen the synchronization features of RN, it would be
interesting to compare them with the performance of Regular
Lattices ($RL$). In the 1D RL we consider, each oscillator is
coupled to its $ 2l/N$ nearest-neighbors in a ring-like (1D
reticle with periodic boundary conditions) network. In Fig. 5
there is an example of $RL$ with $2l/N=4$. In order to do the
comparison we must calculate $T$ for the $RL$, always keeping the
same number of nodes and links as in the RN cases. Since the $RL$
is the topological configuration which is a connected network with
a minimum number of links ($l=N$) we can also explore topologies
with fewer links than the $RN$.

Another point one has to take into account when studying $RL$ with
a growing number of links $l$ is that it is not possible to add
just one link to pass from one configuration to another since it
would break the regularity of the lattice. Instead, one has to
work with integer values of $2l/N$, that is, adding a next nearest
neighbor to {\em all} oscillators when passing from one
configuration to the next one that have more links. Therefore,
altough we can start from an initial minimal configuration with
less links, we have less points to study.

In Fig. 4 results for $\varepsilon=0.01$ are shown. One can
clearly see that the $RL$ performs better than the $RN$ for all
degrees of connectivity. This result holds for all $\varepsilon
>0$. Nevertheless, this difference is only appreciable for lower
values of $l$ so that as our network has more links it quickly
vanishes. When we are close to the globally connected network, the
synchronization features of both kind of networks are roughly the
same, while in the low connectivity case, the $RN$ has
synchronization time $T$ much longer (about twice) than the $RL$.

\begin{figure}
\label{comparacio}
\centerline{
        \epsfxsize= 8.0cm
        \epsffile{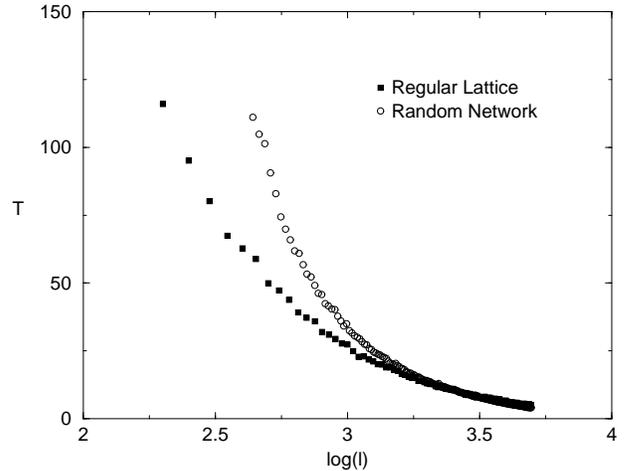}}
        \caption{$T$ as a function of $l$ with a fixed number
        of oscillators $N=100$ and interaction strength
        $\varepsilon=0.01$ for the two kind of extreme networks
        $RL$ and $RN$. $RL$ always performs better than $RN$
        although its difference quickly vanishes as we tend to
        a globally coupled network.}
\end{figure}

\section{Mixed Topologies}
So far, we have checked the synchronization features of the two
extreme kind of networks: RN and the RL. Nevertheless, there
exists a whole family of networks that lie between these two
limits. They are networks of mixed nature, that is, although they
may have some random connections, also posses an underlying
regular structure. Recently, this kind of networks have received a
lot of attention \cite{SW,Nature,barabasi}, specially due to to
the so-called {\em Small-world} networks. These networks,
basically a regular lattice with a very small amount of random
connections, have the advantage of having a low average distance
among nodes while keeping a highly clustered structure. In this
work, we examine synchronization time for networks with all degree
of randomness ranging from the RL to RN.

We parametrically characterize these networks with a {\em
re-wiring} probability per link $p$. It defines the following
randomization procedure: starting from an initial RL of $l$ links,
we {\em cut} each link with probability $p$ and {\em re-wire} it
between two randomly chosen pair of nodes. Notice that our method
slightly differs from other used by some authors who just rewire
one edge of the link \cite{Nature} or add new
ones\cite{percolation}.  In this way, we keep the number of links
$l$ constant and recover the previous two limiting cases, the $RN$
and the $RL$, for $p=1$ and $p=0$ respectively (see Fig. 5).

\begin{figure}
\label{xarxa}
\centerline{
        \epsfxsize= 8.0cm
        \epsffile{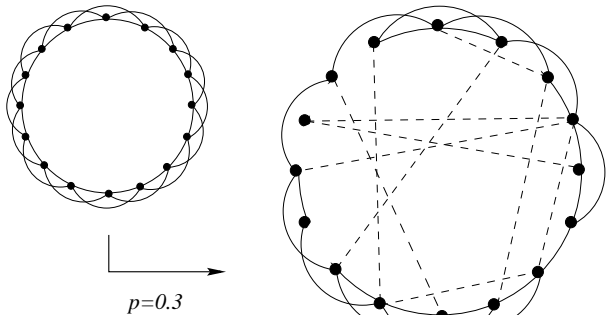}}
        \caption{Randomization procedure for an initial RL with
        links to first and second nearest neighbors. Each links
        is cut with a probability $p=0.3$ and re-wired between
        two randomly selected pair of nodes (dashed lines).
        For $p=0$ we have again the $RL$ since no link is re-wired
        while for $p=1$ the pure RN is recovered. }
\end{figure}

In Fig. 6 we see the synchronization time $T$ for a network of
$N=300$ oscillators with $2l/N=16$. One can clearly see that $T$
grows monotonously as we introduce more disorder into the system
(increasing $p$). For different $N$ and $l$ the behavior of $T$ is
qualitatively the same.

\begin{figure}
\label{T-p}
\centerline{
        \epsfxsize= 8.0cm
        \epsffile{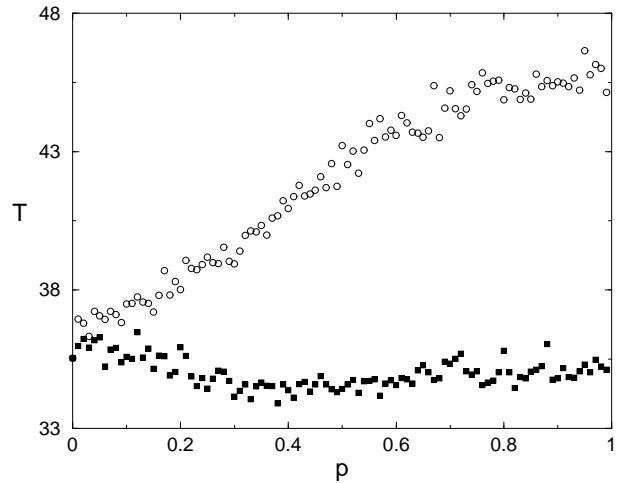}}
        \caption{Synchronization Time $T$ for the whole range of
        re-wiring probability $p$. Each point is averaged over
        1000 realizations for a system of $N=300$ oscillators
        and $2l/N=16$. $T$ increases monotonously with $p$ as
        the dispersion in the number of links also does. This
        figure can be thought as a cross section of Fig. 4, with
        an extra p-axe. Empty circles correspond to the normal case.
        Filled squares show the results once the interactions have
        been normalized according to Eq. \ref{normalitzacio}.}
\end{figure}

These results obviously raise a question: why does topological
disorder slow the synchronization process?. The re-wiring process
induces a random distribution of links for any oscillator.
Therefore, two adjacent units can have a very different number of
oscillators. This fact is crucial since the incoming signal from
the firings of the neighborhood of a given oscillator can be much
larger or smaller than the signal that another of its neighbors
receive. In this case, the two oscillators have different
effective frequencies. The larger the difference in their
effective driving, the more difficult is to synchronize these two
units. This can be thought as a kind of {\em dynamic frustation}
among two adjacent oscillators. One way of quantifying this
problem is to check the variability in the number of neighbours
per oscillator. In Fig. 7 we can see how does the dispersion in
the number of links per node grow as we induce more topological
disorder. This dispersion $\sigma^2$ is zero for $p=0$ ($RL$)
whereas for a $p=1$ ($RN$) the distribution of links is known to
follow a Poisson distribution with a variance equal to $2l/N$ when
$N\rightarrow \infty$ \cite{erdos}. As we can see, both Fig. 6 and
7 look quite similar, they show a monotonic growth with the
re-wiring probability $p$ wich seem to saturate for values close
to $1$.

\begin{figure}
\label{dispersio}
\centerline{
        \epsfxsize= 8.0cm
        \epsffile{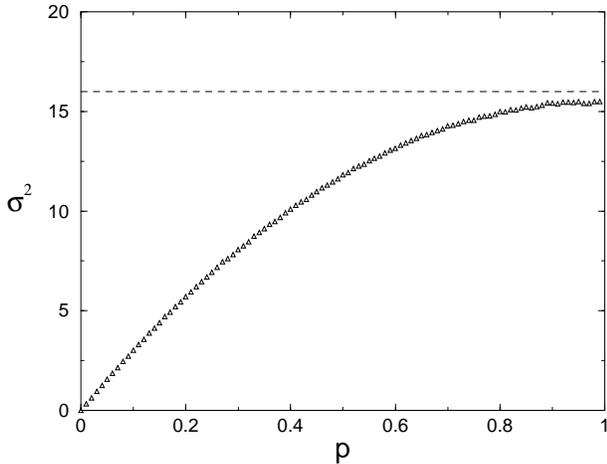}}
        \caption{Dispersion ($\sigma^2$) in the number of links for
        the whole range of p. Each point corresponds to an average
        of 1000 realizations for a system of $N=300$ oscillators
        and $2l/N=16$. For $p=1$ the system has a complete Random
        Network topology and thus is expected to have a
        $\sigma^2\sim l/N$ (Poisson Approximation). Dashed
        line corresponds to this limit behavior}
\end{figure}

Another way to check if, in the topologically disordered model,
this dynamic frustation is the responsible for the delay to
synchronization is trying to remove it. This can be done if we
think in terms of effective drivings, once we have seen that
topological disorder induces an heterogeneity in these drivings,
we can try to make them homogeneous again by means of a convenient
local interaction normalization. The normalization works as
follows, without changing the topology, each oscillator modifies
all pulses it receives from the firing of any of its neighbors by
the factor

\begin{equation}
\label{normalitzacio}
\varepsilon_{i}^{norm}=\varepsilon \frac{<N(\Gamma(i))>}{N(\Gamma(i))},
\end{equation}
where $N(\Gamma(i))$ is the number of neighbors of $\phi_i$. This
normalization means that the more pulses an oscillator receives,
the less intense they are. The average number of neighbours
$<N(\Gamma(i))>$ is always $2l/N$ for all $p$. In Fig. 6 we see
that this procedure does remove the dynamical frustation, lowering
the time needed to achieve synchronization, and even making it
shorter than the unnormalized case for some small values of $p$.
Therefore, with this rough method we are able to get rid of the
effect that topological disorder had on the synchronization
features of the network.

From another point of view, one can think of this variability
induced by the topological disorder as something equivalent to
have some diversity in a population of coupled oscillators on a
$RL$. Imagine that, for instance, a population of oscillators
following the dynamics:

\begin{equation}
\frac{d\phi_i}{dt}=1+\tilde{\varepsilon}_{ij}
\sum_{j \subset \Gamma(i)}\delta(t-t_j),
\end{equation}
with $\tilde{\varepsilon}_{ij}$ being a random variable uniformly
distributed over the interval $(\varepsilon-s,\varepsilon+s)$. In
this case, $s$ gives us a quantitative idea of the population
diversity. Now, in this modified model, synchronization time $T$
also grows as we increase population diversity $s$. In Fig. 8 we
can check this for a population of $N=100$ oscillators in a $RL$
with $2l/N=16$ and a mean value of the interaction
$<\tilde{\varepsilon}_{ij}>=0.01$. The same result, for the
specific case of all-to-all coupling had already been found out in
\cite{bottani}. Therefore, for this kind of pulse-coupled
oscillatory systems, inducing some topological disorder is almost
equivalent to deal with a random distribution of interactions in a
regular lattice as far as synchronization features are concerned.

\begin{figure}
\label{diversity}
\centerline{
        \epsfxsize= 8.0cm
        \epsffile{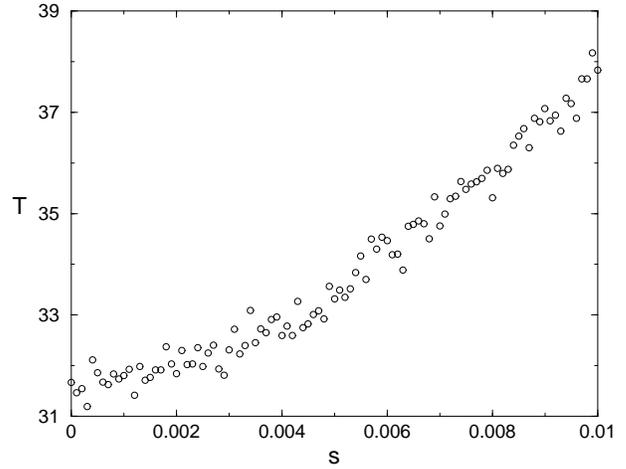}}
        \caption{Synchronization Time $T$ for a population of
        $N=100$ oscillators in a $RL$ with $2l/N=16$ with an
        inhomogeneous distribution of interactions
        $\tilde{\varepsilon}_{ij}$ characterized by its dispersion
        $s$. Increasing dispersion in this distribution makes
        synchronization process more difficult, increasing $T$.
        Results are averaged over different realizations
        of the quenched random interactions and arbitrary
        initial conditions with $<\tilde{\varepsilon}_{ij}>=0.01$.}
\end{figure}

\section{Conclusions}
In this paper we have studied synchronization time $T$ for several
networks, each of them characterized by a different degree of
randomness. For the special case of a completely random network we
have found out a scaling relation between $T$ and network's
connectivity $T(N,l)$. As far as other topologies are concerned,
the regular lattice is the one which synchronizes faster.
Nevertheless, our regular lattice is a 1d ring-like structure, and
there are other kind of regular lattices which might also be
studied (2d lattices, hierarchical trees,...). Therefore the
question of which is the optimal synchronizing network remains
open. However, the main aim of our work was to point out which are
the geometrical mechanisms responsible for slowing or accelerating
the synchronization process in such pulse-coupled systems. It
turns out that the variability in the number of neighbors is a
factor that slows synchronization. We have finally proposed a
local normalization method that manages to remove the effects
induced by the topological disorder. Among the limitations of our
model there is the lack of time delays in the interaction, or a
finite pulse propagation velocity, which are present in real
systems. Such effects might modify some of the results and is part
of future work.

\section*{Acknowledgments}
The authors are indebted to A. Arenas for very fruitful
discussions and to D. J. Watts for sending us a copy of \cite{SW}
prior to its publication. This work has been supported by DGES of
the Spanish Government through Grant No. PB96-0168 and EU TMR
Grant No ERBFMRXCT980183. M. L. acknowledges financial support
from the Spanish Ministerio de Educaci\'{o}n y Cultura, and X.G. from
the Generalitat de Catalunya.

\end{document}